\documentclass[prb,preprint]{revtex4-1} 

\usepackage{amsmath}  
\usepackage{amsfonts} 
\usepackage{graphicx} 
\usepackage{gensymb} 
\usepackage{subfig} 

\begin{document}


\title{Projectile motion: the ``coming and going'' phenomenon}

\author{Williams J. M. Ribeiro}
\email{willribeiro@usp.br} 
\affiliation{Instituto de F\'isica, Universidade de S\~ao Paulo, S\~ao Paulo, Brazil}

\author{J. Ricardo de Sousa}
\email{jsousa@ufam.edu.br}
\affiliation{Departamento de F\'isica, Universidade Federal do Amazonas, Manaus, Brazil}


\date{\today}

\begin{abstract}
An interesting phenomenon that occurs in projectile motion, the ``coming and going'', is analyzed considering linear air resistance force. By performing both approximate and numerical analysis, it is showed how a determined critical angle and an interesting geometrical property of projectiles can change due to variation on the linear air resistance coefficient.
\end{abstract}

\maketitle 

\section{Introduction} 

During introductory courses of elementary physics, many examples of mechanics are studied not considering air resistance effects. We can mention projectile motion, free-fall, harmonic oscillator and many others. Only a brief study of these systems take into account a linear air resistance, which is most studied on classical mechanics courses.\cite{marion} Quadratic air resistance is hardly mentioned, but it can be found in many papers.\cite{ray,belgacem,yabushita,leal} However, in nature air resistance plays a big role on body motion and so it must be included in studying any kinds of phenomena related to mechanics.

Years ago, Walker \cite{walker} worked in an interesting phenomenon related to projectile motion. He showed that, for launch angles bigger than a determined critical angle, the projectile moves away from the origin, approaches it and then moves away again, going against our common sense that a projectile launched should only move away from the origin. Also, he studied a geometrical property displayed by projectiles: the return ellipse. His work was developed considering a projectile motion out of air resistance, in an ideal situation.

The main objective of this paper is to generalize Walker's work, demonstrating if the ``coming and going'' projectile motion also occurs in presence of a linear air resistance force and if there are any changes in the value of the critical angle, mapping how it is related to the air resistance coefficient. We also intend to work on the effects of linear air resistance in the return ellipse.

This paper is organized as follows: in Sec. II we summarize projectile motion and introduce the ``coming and going'' phenomenon in the ideal case; in Sec. III we analyze the effects of air resistance in the ``coming and going'' phenomenon; in Sec. IV we introduce the return ellipse and study the effects of air resistance on it; in Sec. V we draw our conclusions.

\section{Summary of Projectile Motion and the ``Coming and Going'' Phenomenon}

\begin{figure}[h!]
\centering
\includegraphics[width=0.7\textwidth]{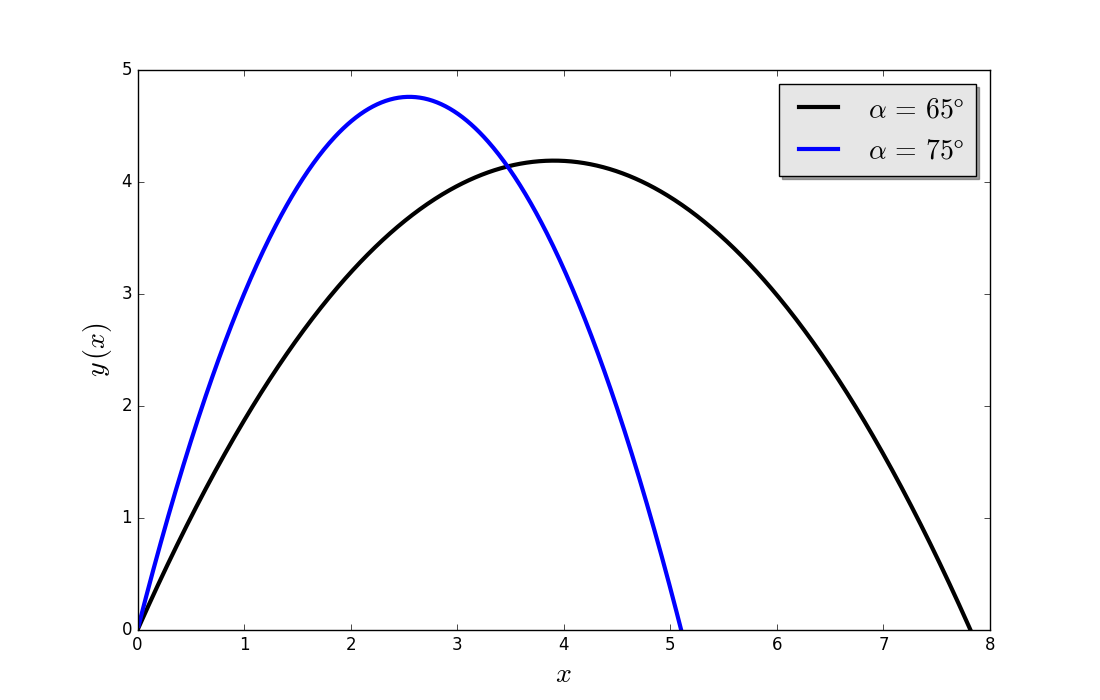}
\caption{Projectiles moving along its trajectories with different launch angles (scales in meters).}
\label{fig1}
\end{figure}

Starting with the equation describing the motion of a projectile with no air resistance,\cite{halliday}

\begin{equation}
\label{eq1}
y(x)=(\tan \alpha)x-\left(\frac{g}{2v_0^2 \cos^2 \alpha}\right)x^2,
\end{equation}
where $\alpha$ is the launch angle, $g$ is the acceleration of gravity and $v_0$ is the launch speed, let's consider a specific situation, without loss of generality, where $v_0=10 \,m/s$, $g=9.8 \,m/s^2$ and with launch angle values $\alpha =75\degree$ and $\alpha =65\degree$. A graphic is displayed in Fig.~\ref{fig1}.

There is a property displayed by the projectile with launch angle $\alpha =75\degree$ not possessed by the one with $\alpha =65\degree$. The second one always moves away from the origin while the first one moves away from the origin, then moves closer and finally moves away again.

In order to analyze this phenomenon, we define the radial distance as the distance between the origin and any point of the p°rojectile's trajectory. Mathematically, the radial distance can be defined as

\begin{equation}
\label{eq2}
r(x) \equiv \sqrt[]{x^2+y(x)^2}.
\end{equation}

\begin{figure}[h!]
\centering
\includegraphics[width=0.7\textwidth]{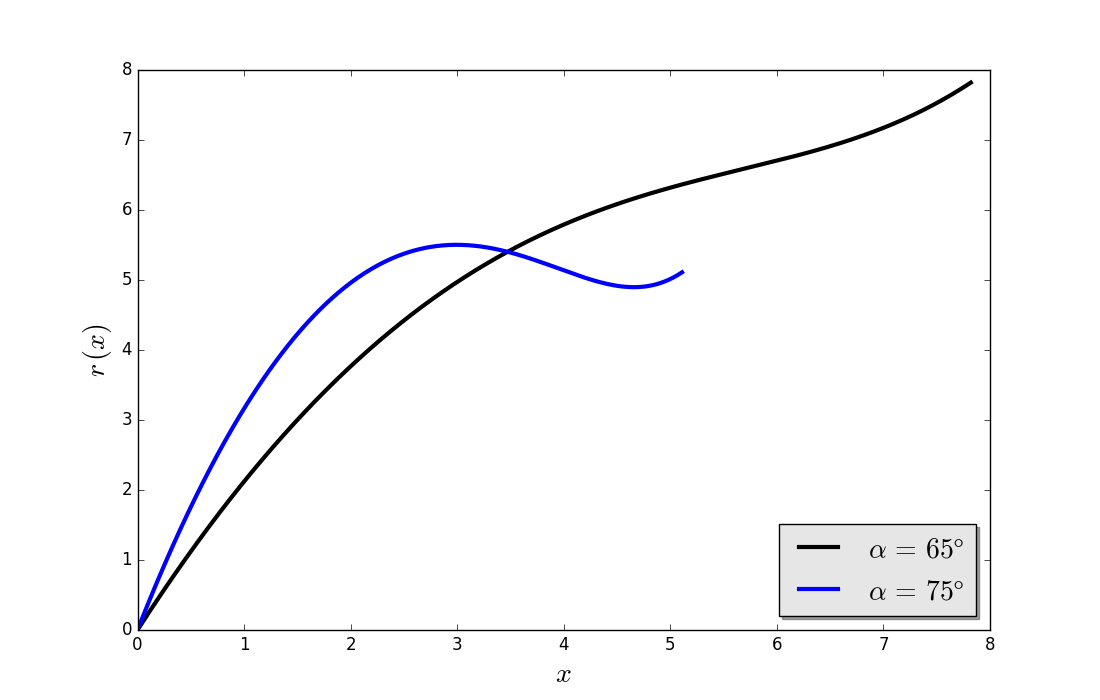}
\caption{Radial distance of the projectiles with different launch angles (scales in meters).}
\label{fig2}
\end{figure}

Now we can see graphically what happens to the radial distance while the projectile is moving if we take a look at Fig.~\ref{fig2}. As we can see, radial distance only increases for the projectile with launch angle $\alpha =65\degree$ while for the projectile with launch angle $\alpha =75\degree$ the radial distance increases, decreases and increases again next to the end of the trajectory. This is a beautiful result, which displays a unique phenomenon.

It can be shown \cite{walker} that there is a critical angle, unique in the ideal situation, from which this phenomenon begins to show up. This critical angle is exactly $\alpha_c = 70.5288... \degree$.

We are now ready to analyze if there is a relation between the critical angle and the air resistance coefficient for linear speed drag force.

\section{``Coming and Going'' Projectile Motion subject to Linear Air Resistance Force}

Considering a drag force in the form $\vec{F_r} = -m\gamma\vec{v}$, the equation for the trajectory of a projectile is given by \cite{stewart}

\begin{equation}
\label{eq3}
y(x) = \left(\frac{g}{\gamma v_0 \cos\alpha}+\tan\alpha\right)x+\frac{g}{\gamma^2}\ln\left(1-\frac{\gamma \sec\alpha}{v_0}x\right),
\end{equation}
where $\gamma$ is the linear air resistance coefficient (experimentally given according to the shape of the projectile) and $m$ is the mass of the projectile. The initial conditions are $x(0)=y(0)=0$, $v_x(0)=v_o \cos\alpha$ and $v_y(0)=v_o \sin\alpha$.

By using dimensionless variables $t^\prime \equiv \gamma t$, $x^\prime \equiv \gamma^2 x/g$, $y^\prime \equiv \gamma^2 y/g$ and $\Gamma \equiv \gamma v_0/g$, we can rewrite Eq.~(\ref{eq3}) as

\begin{equation}
\label{eq4}
y(x) = \left(\frac{\sec \alpha}{\Gamma}+\tan \alpha\right)x + \ln\left(1-\frac{\sec \alpha}{\Gamma}x\right),
\end{equation}
where we maintained our notation without primed variables for convenience. Also, we evaluate the position of the projectile in the dimensionless form as

\begin{equation}
\label{eq5}
x(t) = \Gamma \, \cos\alpha \,(1-e^{-t})
\end{equation}
and

\begin{equation}
\label{eq6}
y(t) = -t + (\Gamma\sin\alpha+1)(1-e^{-t}).
\end{equation}

An attempt to find the way the critical angle varies with the dimensionless resistance coefficient $\Gamma$ using Walker's procedure shows up not to be effective in order to find an analytical solution. However, performing a numerical analysis gives us a good idea of what it happens when linear air resistance is taken into account. By writing the radial distance as a time-dependent variable and taking its derivative, one yields

\begin{equation}
\label{eq7}
\frac{dr}{dt} = \frac{x(t)\dot{x}(t)+y(t)\dot{y}(t)}{\sqrt{x^2(t)+y^2(t)}}.
\end{equation}

\begin{figure}[h!]
\centering
\includegraphics[width=0.7\textwidth]{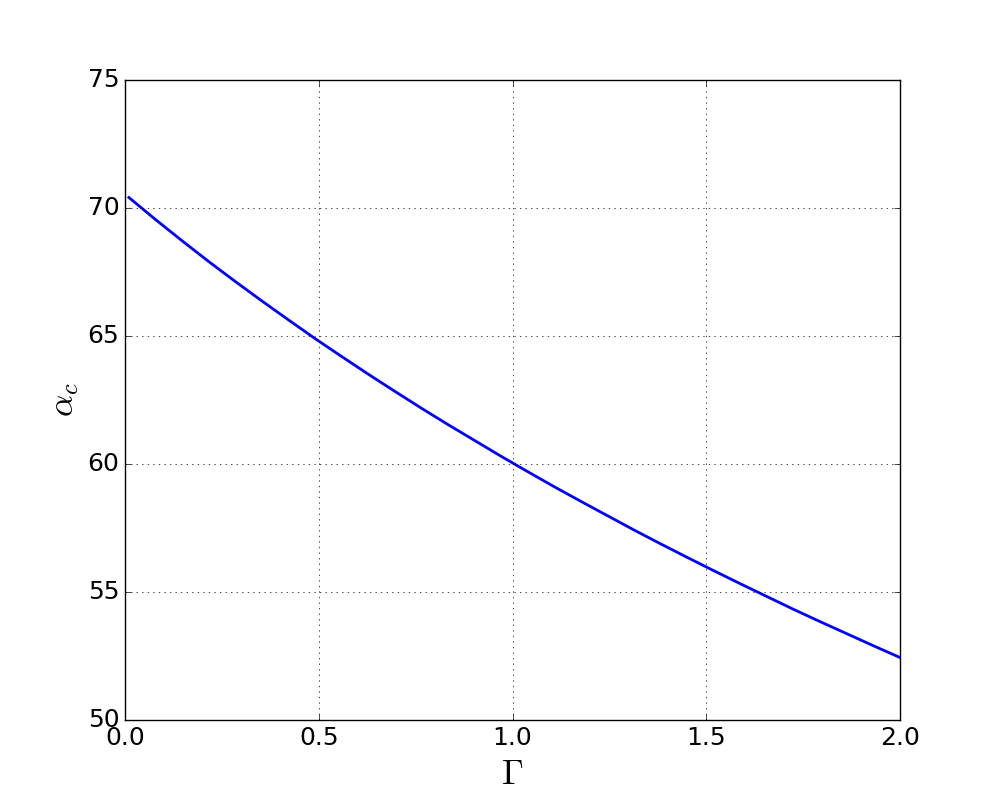}
\caption{Critical angle (in degrees) as a function of the dimensionless air resistance coefficient.}
\label{fig3}
\end{figure}

If the derivative in Eq.~(\ref{eq7}) equals zero, it indicates that for the value chosen for $\Gamma$ the angle $\alpha$ considered in the problem is the critical angle $\alpha_c$. Bearing that in mind, it is easier for the computer to find this limit using a routine, because if the program finds a transition  of $dr/dt$ from positive to negative, it immediately associates the value of $\Gamma$ to the value of the critical angle $\alpha_c$. The evolution of the critical angle is displayed in Fig.~\ref{fig3}.

Although this problem is intractable analytically in its complete form, we can find a solution by considering the approximation $\Gamma<<1$ for small air resistance coefficient. By setting $dr/dt=0$ in Eq.~(\ref{eq7}) together with Taylor expansion, one yields, after some tedious algebra,

\begin{equation}
\label{eq8}
t = \frac{\frac{1}{2}(\Gamma^2+3\Gamma\sin\alpha) \pm \sqrt{\frac{1}{4}(\Gamma^2+3\Gamma\sin\alpha)^2-\frac{2}{3}(\Gamma^4+\Gamma^3\sin\alpha+3\Gamma^2)}}{\frac{1}{3}(\Gamma^2+\Gamma\sin\alpha+3)}.
\end{equation}

\begin{figure}[h!]
\centering
\includegraphics[width=0.7\textwidth]{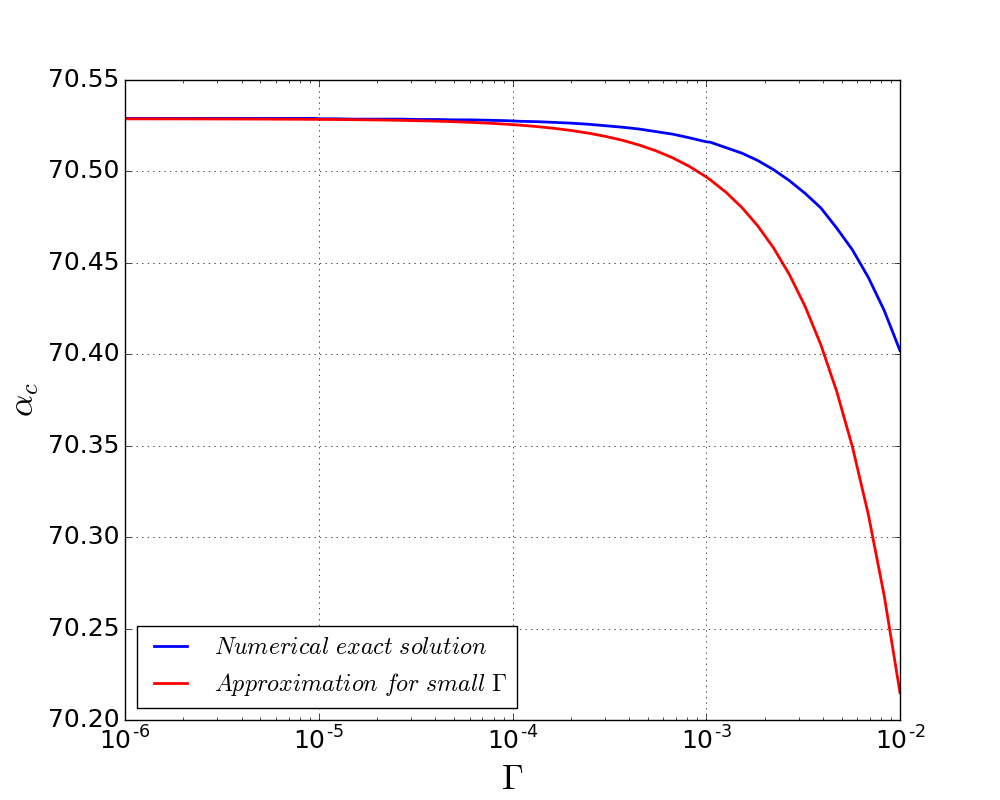}
\caption{Critical angle (in degrees) as a function of the dimensionless air resistance coefficient for $\Gamma<<1$. As we can see, the numerical exact solution matches the analytical approximate solution when $\Gamma$ gets close to zero, as expected.}
\label{fig4}
\end{figure}

Real solutions begin at the angle where the term in the square root is equal to zero, in which one obtains

\begin{equation}
\label{eq9}
\alpha_c(\Gamma) = \sin^{-1}\left(-\frac{5}{27}\Gamma+\frac{2}{9}\sqrt{\frac{40}{9}\Gamma^2+18}\right),
\end{equation}
and the graph is displayed in Fig.~\ref{fig4}.

\section{The Return Ellipse}

Besides the analysis of the critical angle, Walker also analyzed an interesting geometrical property of projectiles related to the critical angle: the return ellipse. This one is defined as the locus of points for which $dr/dt=0$. In the ideal case, Walker showed that this locus of points corresponds to an ellipse centered at $x=0$ and $y=v_0^2/4g$ and that the physical significance of the ellipse is that projectiles moving with trajectories crossing the ellipse when moving toward the ground $(dy/dt<0)$ are also moving toward the origin $(dr/dt<0)$, which means that for these projectiles the launch angle is bigger than the critical angle.

\begin{figure}[h!]
\centering
\includegraphics[width=1.1\textwidth]{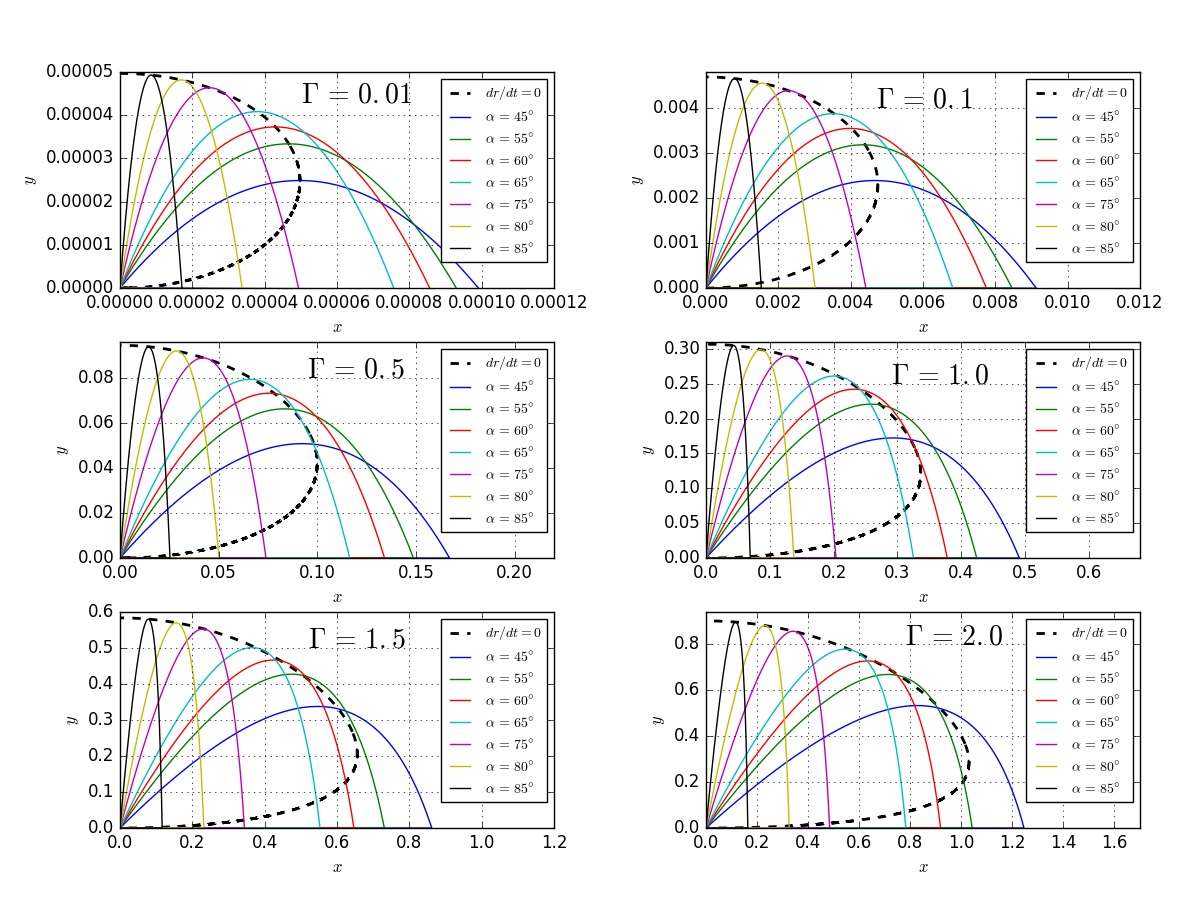}
\caption{Return ellipses (dashed) for different dimensionless resistance coefficients along with projectile trajectories ($x$ and $y$ are dimensionless coordinates).}
\label{fig5}
\end{figure}

In order to see what happens to the return ellipse when linear air resistance is taken into account, we plotted in Fig.~\ref{fig5} the return ellipses obtained numerically for six values of $\Gamma$ along with trajectories of projectiles for different launch angles. We obtained these return ellipses by evolving Eq.~(\ref{eq7}) and saving the positions at the transition points in which $dr/dt$ changes its sign (which are two times for projectiles with launch angles bigger than the critical angle).

As expected, for $\Gamma=0.01$ and $\Gamma=0.1$ projectiles with launch angles $\alpha=85\degree$, $\alpha=80\degree$ and $\alpha=75\degree$ are the only ones moving toward the ground passing inside the return ellipse. However, for $\Gamma=0.5$ projectiles with smaller launch angles (below the critical angle in the ideal case) begin to enter the return ellipse! The new projectile falling to the ground inside the ellipse is the one with launch angle $\alpha=65\degree$. This totally agrees with Fig.~\ref{fig3} in which we see that the critical angle for $\Gamma=0.5$ is smaller than $65\degree$. The subsequent graphics also agree with Fig.~\ref{fig3}. For example, we see that for $\Gamma=1.0$ the projectile with launch angle $\alpha=60\degree$ is entering the ellipse, for $\Gamma=1.5$ the projectile with launch angle $\alpha=55\degree$ is about to enter the ellipse and for $\Gamma=2.0$ the projectile with launch angle $\alpha=55\degree$ has already entered the ellipse.

Walker also demonstrated in the ideal case that the locus of points corresponding to $dr/dt=0$ is coincidently the same locus of points corresponding to $dy/dt=0$ (maximum height), fact which he stated as ``rather unexpected'' because these two requirements correspond to very different conditions. Surprisingly, when we take into account air resistance, this property is not satisfied anymore! In order to see this, for plotting the locus of points for $dy/dt=0$ let's use the equations that define the dimensionless maximum height of the projectile with air resistance,

\begin{equation}
x_{max} = \frac{1}{2}\,\frac{\Gamma^2\sin(2\alpha)}{1+\Gamma\sin\alpha}
\end{equation}
and 

\begin{equation}
y_{max} = \Gamma\sin\alpha - \ln(1+\Gamma\sin\alpha).
\end{equation}

\begin{figure}[h!]
\centering
\includegraphics[width=1.1\textwidth]{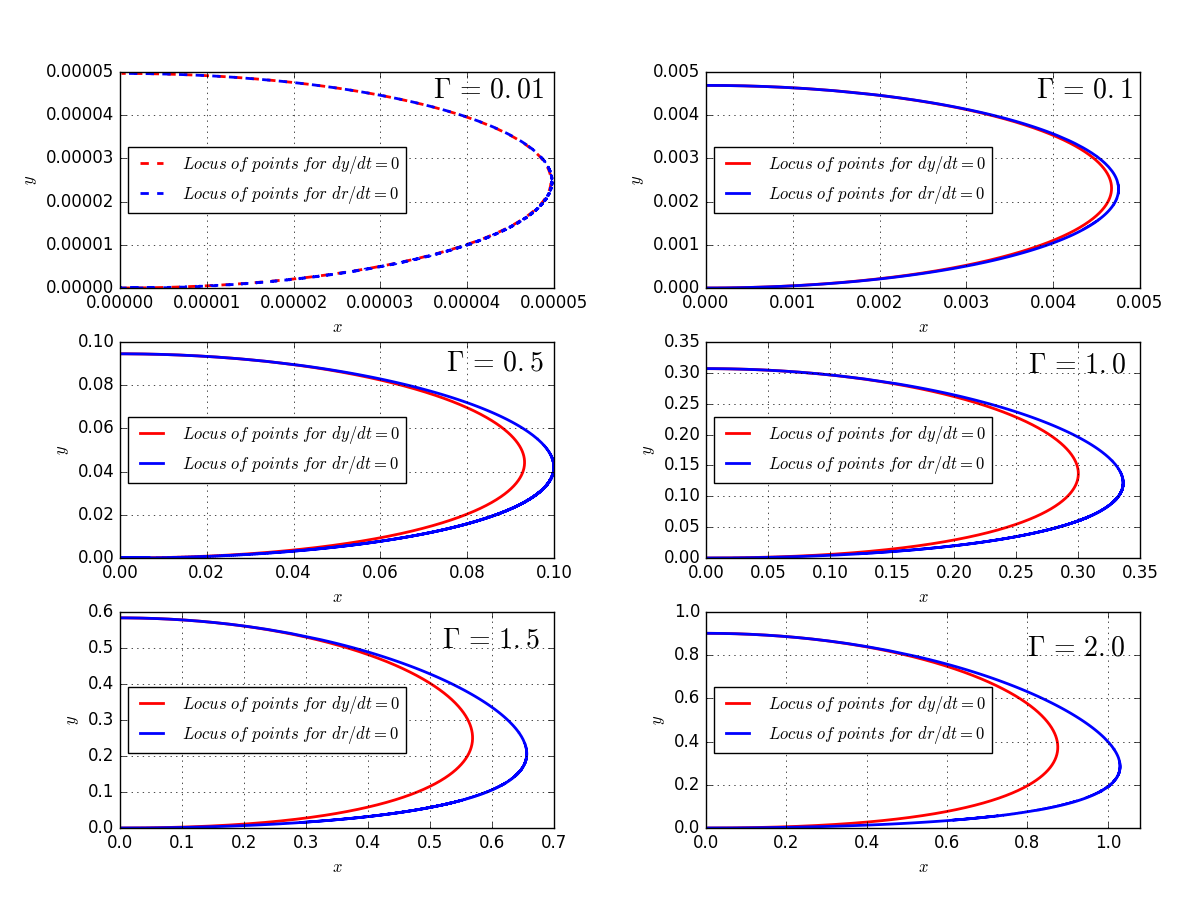}
\caption{Return ellipses (blue) for different dimensionless resistance coefficients along with loci of points for maximum height (red) of projectiles ($x$ and $y$ are dimensionless coordinates).}
\label{fig6}
\end{figure}

The plots are displayed in Fig.~\ref{fig6}. As we can see, for $\Gamma=0.01$ the two curves still agree with each other, but when we go to higher values of $\Gamma$ the two curves don't match each other anymore and the difference between then increases more and more. Therefore, the equality between $dr/dt=0$ and $dy/dt=0$ is only valid for the ideal case $\Gamma=0$.

\begin{figure}[h!]
\centering
\includegraphics[width=1.1\textwidth]{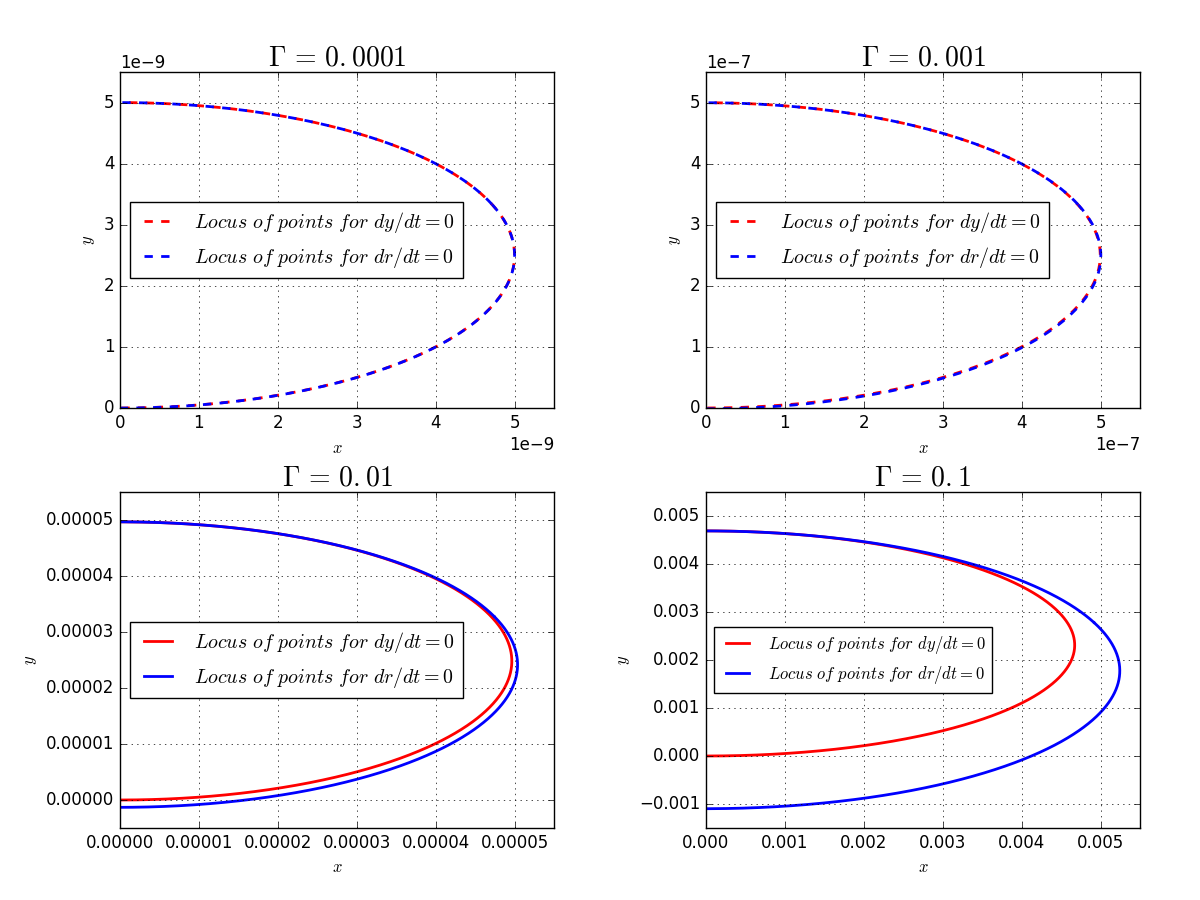}
\caption{Approximate return ellipses (blue) for different dimensionless resistance coefficients along with loci of points for maximum height (red) of projectiles ($x$ and $y$ are dimensionless coordinates).}
\label{fig7}
\end{figure}

Just for checking, it is interesting to see if our approximate solution (\ref{eq8}) is still valid for the construction of the locus of points for $dr/dt=0$ at low $\Gamma$. By substituting Eq.~(\ref{eq8}) into Eq.~(\ref{eq5}) and Eq.~(\ref{eq6}), one finds the plots displayed in Fig.~\ref{fig7}. For small values of $\Gamma$ ($0.0001$ and $0.001$) we don't see visible differences between the curves, while for $\Gamma=0.01$ we can see already that our approximate solution is not valid anymore.

\section{Conclusions}

The aim of this paper was to check on how an interesting phenomenon that occurs in projection motion, the ``coming and going'', is affected by a linear air resistance force. First we summarized how this phenomenon occurs for an ideal situation, out of air resistance, and then showed how a linear air resistance affects the phenomenon.

When we analyze the ideal situation, it can be demonstrated that the critical angle is unique and does not change even if other variables are changed, such as the initial velocity of the projectile. Including air resistance, it was possible to show that the critical angle for the phenomenon is not unique anymore, but it changes with linear air resistance coefficient.

Figure~\ref{fig3} shows that the curve relating the critical angle with dimensionless air resistance coefficient is decreasing. This is an expected result, because with an increasing air resistance coefficient the projectile suffers an increasing resistance force, reducing the value of the critical angle. Another interesting observation that comes from Fig.~\ref{fig4} is that for a null resistance coefficient the critical angle tends to approximately $\alpha=70.528...\degree$, the exact result found theoretically in ideal projectile motion, further confirming the numerical analysis.

Our approximate approach turned out to be effective for three reasons: when considering $\Gamma=0$ case in Eq.~(\ref{eq9}) we recover the result for the ideal case for $\alpha_c$; Fig.~\ref{fig4} shows that the approximate and numerical solutions match in the limit $\Gamma<<1$; Fig.~\ref{fig7} shows that the approximate and analytical solutions also match in the limit $\Gamma<<1$.

We also showed that the matching between the return ellipse and the locus of points that define the maximum height of projectiles doesn't exist anymore when air resistance is accounted for. This is a beautiful result that certainly is not trivial at first.

As an extension of this work, it would be interesting to consider more realistic situations as quadratic drag, lift, wind, etc, retaining a numerical approach in order to analyze these cases.

\end{document}